\documentclass[journal,onecolumn]{IEEEtran}

\usepackage{booktabs, balance}
\usepackage{dcolumn,multirow}
\usepackage{textcomp}
\usepackage[utf8]{inputenc}
\usepackage[T1]{fontenc}
\usepackage[T1]{fontenc}
\usepackage{hyperref}
\usepackage{array}
\usepackage{listings}
\usepackage[sort, numbers]{natbib}
\usepackage{verbatim}
\usepackage{wrapfig}
\usepackage{svg}
\usepackage{caption}
\usepackage{subcaption}

\usepackage{multicol}

\usepackage{float}

\newcommand{\ignore}[1]{}

\newcolumntype{P}[1]{>{\centering\arraybackslash}p{#1}}

\newcolumntype{L}[1]{>{\raggedright\let\newline\\\arraybackslash\hspace{0pt}}m{#1}}
\newcolumntype{C}[1]{>{\centering\let\newline\\\arraybackslash\hspace{0pt}}m{#1}}
\newcolumntype{R}[1]{>{\raggedleft\let\newline\\\arraybackslash\hspace{0pt}}m{#1}}

\definecolor{dkgreen}{rgb}{0,0.6,0}
\definecolor{gray}{rgb}{0.5,0.5,0.5}
\definecolor{mauve}{rgb}{0.58,0,0.82}

\AtBeginDocument{%
  \providecommand\BibTeX{{%
    \normalfont B\kern-0.5em{\scshape i\kern-0.25em b}\kern-0.8em\TeX}}}

\begin{document}


\title{Leveraging LLM Tutoring Systems for Non-Native English Speakers in Introductory CS Courses}







\author{
    \IEEEauthorblockN{
        Ismael Villegas Molina\IEEEauthorrefmark{1},
        Audria Montalvo\IEEEauthorrefmark{1},
        Benjamin Ochoa\IEEEauthorrefmark{1},
        Paul Denny\IEEEauthorrefmark{2},
        Leo Porter\IEEEauthorrefmark{1}
    } \\
    \IEEEauthorblockA{\IEEEauthorrefmark{1} University of California, San Diego \\
    \{isvilleg, ansaravi, bochoa, leporter\}@ucsd.edu} \\
    \IEEEauthorblockA{\IEEEauthorrefmark{2} University of Auckland \\
    paul@cs.auckland.ac.nz}
}

\maketitle

\begin{abstract}


Computer science has historically presented barriers for non-native English speaking (NNES) students, often due to language and terminology challenges. With the rise of large language models (LLMs), there is potential to leverage this technology to support NNES students more effectively. Recent implementations of LLMs as tutors in classrooms have shown promising results. In this study, we deployed an LLM tutor in an accelerated introductory computing course to evaluate its effectiveness specifically for NNES students. Key insights for LLM tutor use are as follows: NNES students signed up for the LLM tutor at a similar rate to native English speakers (NES); NNES students used the system at a lower rate than NES students---to a small effect; NNES students asked significantly more questions in languages other than English compared to NES students, with many of the questions being multilingual by incorporating English programming keywords. Results for views of the LLM tutor are as follows: both NNES and NES students appreciated the LLM tutor for its accessibility, conversational style, and the guardrails put in place to guide users to answers rather than directly providing solutions; NNES students highlighted its approachability as they did not need to communicate in perfect English; NNES students rated help-seeking preferences of online resources higher than NES students; Many NNES students were unfamiliar with computing terminology in their native languages. These results suggest that LLM tutors can be a valuable resource for NNES students in computing, providing tailored support that enhances their learning experience and overcomes language barriers.
\end{abstract}



\section{Introduction}
\label{sec:introduction}




In computing education, being a non-native English speaker (NNES) can present learning challenges, especially for those outside anglophone countries or with limited English proficiency, due to the predominance of English-based materials~\cite{becker2019parlez}. 
The most popular programming languages (e.g., Java, Python, C) use English for their keywords and API naming conventions~\cite{hanselman2008you, chistyakov2017language, igawa2017non}. English computer terminology used in classroom settings is one of the main factors negatively affecting NNES student performance in programming classes~\cite{keen2009predicting}, predicting approximately 50\% of final exam performance when controlling for English proficiency~\cite{alaofi2022validated}. Moreover, English proficiency is strongly correlated with programming performance~\cite{qian2016correlates, rauchas2006language, wibowo2022correlation, pudyastuti2014correlation}, further widening the gap for NNES students. In addition to programming languages being primarily based in English, most online resources for learning programming (e.g., documentation, online Q\&A forums, tutorials) are also predominantly in English. These skewed linguistic accommodations make a certain level of English proficiency increasingly necessary to learn computer programming~\cite{hanselman2008you}. This de facto prerequisite 
creates a significant hurdle for NNES individuals who may have limited English proficiency~\cite{guo2018non}. While tailoring the delivery of materials for NNES students has been suggested~\cite{becker2022horses}, the practical implementation of this strategy has traditionally been a challenge. However, large language models (LLMs) may now provide an opportunity to better serving NNES computing students as the models having been trained on 100+ natural languages~\cite{xue2020mt5}. Indeed, with the multilingual capabilities of LLMs and their past implementation in computing classrooms~\cite{denny2024computing, prather2023robots, liu2024beyond, liu2024Teaching, woodrow2024ai}, this technology might help alleviate the de facto primarily English setting by allowing NNES students to easily obtain assistance in their native language.


With the rapid rise of LLMs, computing educators face the challenge of integrating this technology into the classroom~\cite{becker2023programming}. The study by \citeauthor{denny2023can} suggests that LLMs can enhance the accessibility and scalability of high-quality educational content while reducing educators' workloads~\cite{denny2023can}. One implementation of this technology is through tutoring systems that provide guidance, rather than direct answers to programming questions, helping users to reach solutions independently~\cite{kazemitabaar2024codeaid, liffiton2023codehelp}.

LLM tutoring systems have shown positive effects for students in the classroom~\cite{kazemitabaar2024codeaid, liffiton2023codehelp, zapatadesarrollo}. Coupled with this, state-of-the-art LLMs that have been trained on 100+ languages~\cite{xue2020mt5} can perform extremely well when translating between languages~\cite{jordan2024need}. Due to this, we hypothesize that NNES students may especially benefit from the use of LLM tutoring systems to ask for help---including in their native language. To understand this, we asked the following research questions:
\begin{enumerate}
    \item How do NNES students interact with a LLM tutor, and how does this differ from native English speaker (NES) students?
    \item How do NNES students' views differ from NES students in regards to LLM tutor use?
\end{enumerate}

\section{Related Work}
\label{sec:background}




\subsection{Non-Native English Speaker Stressors and LLM Assistance}
Past research has highlighted several challenges faced by NNES students in academic environments, particularly in fields like computer science~\cite{keen2009predicting, hanselman2008you, guo2018non, alaofi2023use}. \citeauthor{alaofi2023use} highlighted various NNES student difficulties in CS1, including reading, understanding, or using error messages, understanding or using hints and tips from Integrated Development Environments in English, and general English language difficulties~\cite{alaofi2023use}. Due to a mixture of these difficulties, NNES students often experience linguistic anxiety~\cite{pappamihiel2001moving} and communication apprehension~\cite{sobotka2020role} which can be exacerbated in an academic setting where technical terminology is prevalent. These hurdles, coupled with the embarrassment that NNES students often feel when asking for help~\cite{agarwal2022analysis}, can impact their learning process and willingness to seek necessary assistance. Additionally, tutors have reported additional time required to support NNES students~\cite{hope2003international}, which is in line with tutors reporting overall difficulty in implementing best tutor practices due to time constraints~\cite{riese:ta:challenges}.

\subsection{LLMs in the Computing Classroom}
\label{subsec:background:llm-in-computing}

The increasing use of LLMs in computing classrooms~\cite{denny2024computing, prather2023robots, liu2024beyond, liu2024Teaching, denny2024explain, woodrow2024ai, denny2024prompt} presents a valuable resource for students and new opportunities to support learning. State of the art models provide an alternative avenue for help seeking and can therefore aid learning while potentially reducing the workload on tutors~\cite{denny2024desirable, liu2024Teaching}, although potential challenges to learning also exist \cite{prather2024widening}.  Recent work also points to particular benefits for NNES students \cite{smith2024explainplainlanguagequestions}. 
Within computing education, NNES students have been found to grasp computing concepts more quickly when using their native language~\cite{dasgupta2017learning}.
Given the strong performance of LLMs in non-English languages~\cite{whitehouse2023llm, jiao2023chatgpt, holtermann2024evaluating} and the fact that many models are trained on over 100 languages~\cite{xue2020mt5}, LLMs could be particularly beneficial for students studying in their preferred language. LLMs have the potential to effectively bridge language gaps in computing education, which is especially important given that learning to interact productively with LLMs is likely to become an essential skill for all students~\cite{denny2023conversing}.

This leads to our study in which we use the CodeHelp LLM tutor~\cite{liffiton2023codehelp, sheese2024patterns}. CodeHelp is an open-source LLM-powered tool designed with guardrails to offer on-demand assistance to computing students without directly providing code solutions. We chose this tool as prior research indicated that it was well-received by students and instructors alike, and presented users with a simple interface~\cite{liffiton2023codehelp}. CodeHelp was praised for its availability, help with resolving programming errors, easy deployment, and its supplemental support~\cite{liffiton2023codehelp}. 

\section{Method}
\label{sec:method}
\subsection{LLM Tutor Deployment and Study Context}

\begin{figure}
    \centering
    \caption{CodeHelp user interface illustrating input fields}
    \includegraphics[width=0.65\linewidth]{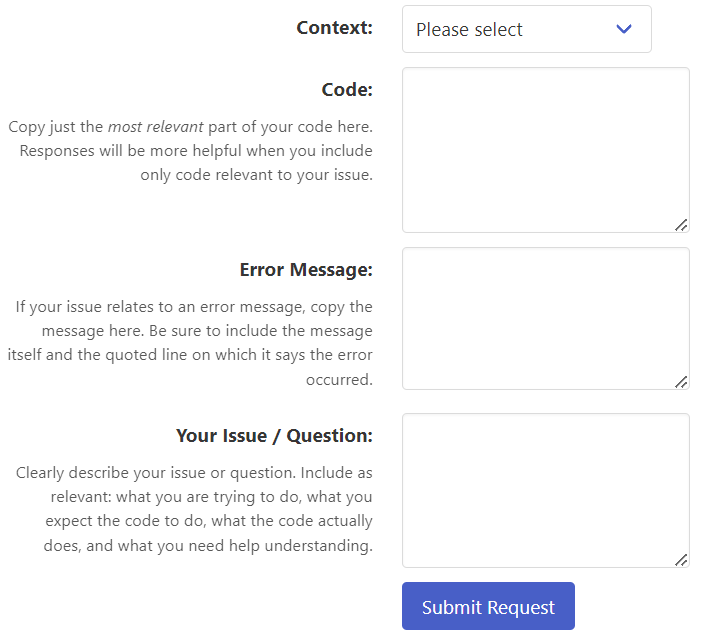}
    \label{fig:codehelp-system}
\end{figure}

In this study, we use the CodeHelp system---an open source LLM-powered tutor that includes built-in guardrails to produce general guidance toward a solution to a programming problem while being prevented from providing direct solutions~\cite{liffiton2023codehelp}. Students interact with CodeHelp in a simple form, providing the programming language of interest (e.g., Java), an extract of the code related to the doubt, the error message, and the student's question. The form can be seen in Figure~\ref{fig:codehelp-system}. The written question is what will be analyzed in this study.

The implementation of CodeHelp aimed to enhance students' learning experiences by providing an accessible, AI-driven resource for supplementary instruction and problem-solving support. CodeHelp leverages the OpenAI API with a choice to use the GPT-3.5-Turbo or the GPT-4o models. In our study, we used the GPT-3.5-Turbo model for financial reasons. CodeHelp was introduced to the course via a class forum post from the instructor. After some use with the system, we noticed that the CodeHelp system would always provide help in English, regardless of the language of the query. Due to the nature of this study, we coordinated with the CodeHelp developer to update the system to reply with the same natural language in which a question was asked, a characteristic known as language fidelity~\cite{holtermann2024evaluatingelementarymultilingualcapabilities}.

The study took place in an accelerated introductory computer science course at a large research-focused institution in North America. This course is designed to provide students with a comprehensive foundation in computer science principles within a condensed timeframe for students with prior programming experience. The course used the Java programming language and topics covered are variables, conditionals, loops, functions, class design, inheritance, polymorphism. In-person tutor office hours were offered on weekdays from 9AM to 9PM with the exception of during lecture and discussion times. Online tutor office hours were offered on weekends from 10AM to 5PM. All assignments were due by Wednesday at 11:59PM local time. Lastly, the average response time for student questions on the class forum (Piazza) was 20 minutes.

Participants for this study were recruited from the pool of 321 students enrolled in the aforementioned accelerated course. All students in the course had access to the CodeHelp tool, but were not required to participate in our study surveys. CodeHelp was available to students for a total of five weeks. Each student independently signed up to the CodeHelp system, ensuring individualized tracking of interactions and usage patterns. Enrollment in the study was voluntary, and students were informed of the study's purpose, procedures, and potential benefits prior to participating. In order to incentivize survey completion, the professor of the course offered extra credit points for completing  surveys but students could still chose to participate in the study or not. In total, 80 students participated in the first survey and 162 students participated in the second survey, for a total of 170 unique students. Of these 170 students, 109 were native English speakers and 61 were non-native English speakers (as seen on Table~\ref{tab:codehelp-usage-results}). All participation procedures were in accordance with our approved human subjects protocol.

\subsection{Survey Design and Administration}
\label{subsec:methods:survey-design-administration}
To assess the impact of the LLM tutor on students' learning experiences, a two-phase survey approach was employed. After four weeks of using the system, participants were asked to complete an initial survey. This survey included open-ended questions designed to evaluate their perceptions of the LLM tutor's effectiveness, ease of use, help-seeking preferences, and overall satisfaction. It also gathered qualitative feedback on specific features and areas for improvement.

Following the analysis of the first survey's open-ended question responses, the research team identified key themes and areas requiring further investigation, which informed the design of the more targeted second survey. This follow-up survey included specific questions that probed deeper into themes from the initial survey, allowing the team to collect more detailed and focused data on emerging issues. This second survey was administered during the final week of the course. 

\subsection{Data Analysis}

CodeHelp use and survey responses were anonymized and analyzed.

To answer RQ1 around interactions with the LLM tutor, the enrollment and usage rates were analyzed using a $\chi^{2}$ test to explore differences between NNES and NES students. Specifically, we compared the proportions of NNES and NES students who enrolled in CodeHelp and those who actively used the platform post-enrollment. Additionally, we calculated the unique users by hour of a given day to analyze usage patterns for each group. For natural language use, the $langdetect$ Python package was used to identify the languages asked. A manual inspection of all non-English queries was also carried out to verify the correct language identification. We examined the frequency and provide descriptive statistics on the languages used to ask queries to CodeHelp.

To answer RQ2, regarding perceptions towards LLM tutor use, survey responses for open-ended and Likert questions were analyzed. For open-ended responses, the first two authors of this paper conducted an inductive thematic analysis to explore emerging themes of opinions from participants. All quotes from participants were considered. As the team independently reviewed and coded the data, each researcher noted patterns and developed initial themes. The team then reconvened to compare results, where disagreements were noted for each theme description and corresponding quotes. When disagreements arose, through the process of negotiated agreement the team collaboratively refined their themes -- ultimately reaching a consensus that captured the participants' experiences authentically. For Likert questions, independent samples t-tests were used to identify differences in opinions between NNES and NES students.

\section{Results}
\label{sec:results}

\subsection{RQ1: LLM Tutor Usage}
\label{subsec:results:llm-usage}

The LLM tutor use was evaluated with the following: CodeHelp enrollment rate, CodeHelp usage rate after enrollment, unique users per day per hour, and natural language use. The NNES designation was self-reported by students through the surveys. As a result, if a student did not participate in either survey we did not have access to their NNES designation so their CodeHelp data was dropped from the study. If a student had conflicting NNES designation from one survey to another due to minor differences in how we phrased the NNES question, their latest designation (survey 2) was used for the CodeHelp data.

\subsubsection{\textbf{CodeHelp Enrollment and Usage}}
\label{subsubsec:results:llm-usage:enrollment-and-usage}

\begin{table}
    \centering
    \caption{CodeHelp Enrollment and Usage Rates for NNES and NES Students}
    \label{tab:codehelp-usage-results}
    \begin{tabular}{L{2.75cm}|C{1.25cm}|C{1.25cm}|C{1.25cm}}
		\toprule
		\hline
		 & NNES & NES & Total \\
		\hline
		Total Class Makeup & 61 & 109 & 170 \\
		CodeHelp Enrolled & 37 & 69 & 106 \\
		Queries Asked & 269 & 674 & 943 \\ 
        \hline
        \hline
        Enrollment Rate & 60.7\% & 63.3\% & 62.4\% \\
        Queries Per Student & 7.3 & 9.8 & 8.9 \\
		\bottomrule
    \end{tabular}
\end{table}

Data on student enrollment and volume of queries asked can be found in Table~\ref{tab:codehelp-usage-results}.

Table~\ref{tab:codehelp-usage-results} 
presents the enrollment rates for each group. To examine differences in student enrollment, we conducted a $\chi^{2}$ test comparing the overall class composition to the number of students enrolled with the CodeHelp system. The analysis yielded a $\chi^{2}$ statistic of 0.04 with a $p$-value greater than 0.05, indicating no significant difference in enrollment rates.

Table~\ref{tab:codehelp-usage-results} 
also presents the usage rate (queries per student) for each group. To assess differences in CodeHelp use, a $\chi^{2}$ test was conducted by comparing the number of students enrolled in the CodeHelp system to the number of queries made. The analysis yielded a $\chi^{2}$ statistic of 16.89 with $p \ll 0.01$, indicating a significant difference in the number of queries made with NNES students making fewer queries. An effect size of $\varphi = 0.13$ was calculated, suggesting a small effect.

\subsubsection{\textbf{Unique Users per Day per Hour}}
\label{subsubsec:results:llm-usage:codehelp-usage}
\begin{figure*}
    \centering

    \caption{Unique Users per Day per Hour}


    \begin{subfigure}{0.45\textwidth}
        \centering
        \caption{Native English Speakers}
        \includegraphics[width=\linewidth]{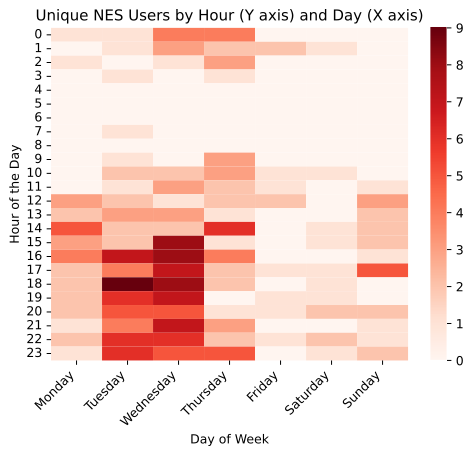}
        \label{fig:unique-nes-query-time}
    \end{subfigure}  
    \begin{subfigure}{0.45\textwidth}
        \centering
        \caption{Non-native English Speakers}
        \includegraphics[width=\linewidth]{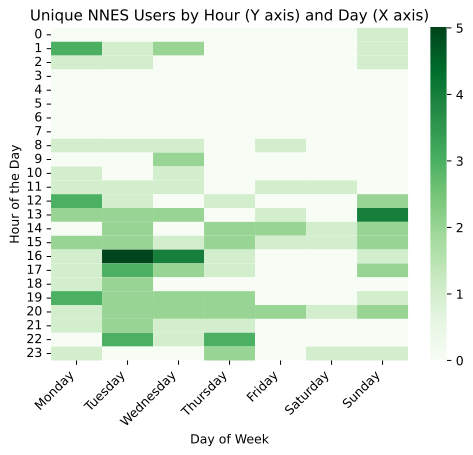}
        \label{fig:unique-nnes-query-time}
    \end{subfigure}

    \label{fig:unique-users-query-time}

\end{figure*}

Figure~\ref{fig:unique-users-query-time} illustrates the number of unique users submitting queries for a given day in a given hour. We can see that NNES and NES students show different patterns of CodeHelp use. NES students typically used CodeHelp on Tuesday and Wednesday evenings before the deadline. NNES students typically used CodeHelp on Tuesday evenings with a more even spread across the week.

\subsubsection{\textbf{Language Use}}
\label{subsubsec:results:llm-usage:languages}
\begin{table}
    \centering
    \caption{Non-English Languages Used in Queries}
    \vspace{-8pt}
    \label{tab:query-languages}
    \begin{tabular}{L{3.75cm}|C{1.5cm}|C{1.5cm}}
		\toprule
		\hline
		\textbf{Language (ISO 639-1 Code)} & \textbf{NNES} & \textbf{NES} \\
		\hline
		Chinese (zh) & 17 & 0 \\ \hline
		Hindi (hi) & 3 & 1 \\ \hline
		Korean (ko) & 3 & 0 \\ \hline
		Arabic (ar) & 1 & 0 \\ \hline
		French (fr) & 1 & 0 \\ \hline
		Spanish (es) & 0 & 1 \\ \hline
		Vietnamese (vi) & 0 & 1 \\ \hline
        \hline
        \textbf{Monolingual} & 8 & 1 \\ \hline
        \textbf{Multilingual} & 17 & 2 \\ \hline
		\bottomrule
    \end{tabular}
    \vspace{-8pt}
\end{table}

\begin{figure}
    \centering
    \caption{Non-English Query Examples}
    \includegraphics[width=0.75\linewidth]{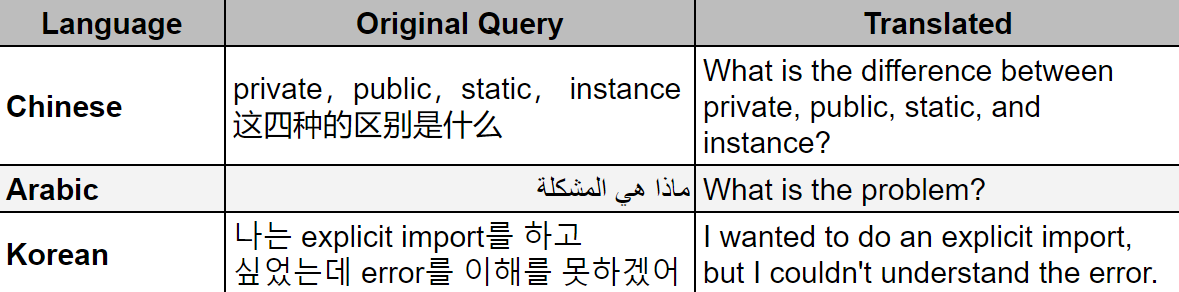}
    \label{fig:non-English_queries}
\end{figure}

The results of the languages used can be found in Table~\ref{tab:query-languages}. From the 269 total queries from NNES students (Table~\ref{tab:codehelp-usage-results}), 25 of them were non-English, resulting in 9\% of NNES queries being non-English. Additionally, we see NES students ask a few non-English queries as well. We note that many of the queries were multilingual in nature with English primarily being used for Java keywords. Figure~\ref{fig:non-English_queries} shows examples of the non-English queries that were submitted by students.

\subsection{RQ2: LLM Tutor Views}
\label{subsec:results:llm-views}

\subsubsection{\textbf{General Views}}
The first survey posed open-ended questions to participants which were thematically analyzed by the research team. We now present seven themes that emerged from the survey along with representative quotes from various NNES and NES students that provide context for each theme.

\paragraph{CodeHelp is More Accessible than Tutors}

Students found the CodeHelp system to be a valuable resource due to its accessibility compared to tutors. \textit{``Because when I need to ask TA, I need to wait as queue and spend lots of time on waiting and communicating. However, Codehelp can solve my 80\% problems just in ten seconds! Furthermore, as a non-native English, I can just ask the question in my comfortable way.'' -- NNES}

More specifically, NNES students found CodeHelp approachable due to not needing to communicate in perfect English. \textit{``As stated before, I struggle with English, it's easier to type than to speak. Thus, making it easier to work with CodeHelp without any waiting nor confusion due to my poor speech.'' -- NNES}

\paragraph{CodeHelp was Instructor Approved}
\label{subsubsubsec:results:llm-views:general:instructor-approved}
Students felt that that since CodeHelp was provided by the instructor, it removed ambiguity on the use of a specific LLM resource. \textit{``I like that it is an approved method of getting help for my CS classes.'' -- NES}

\paragraph{CodeHelp is Conversational}
Natural language communication with CodeHelp was beneficial for students compared to their typical use of other online resources. \textit{``I didn't need to use keywords as if I was just looking up what I needed on google. I could just use natural language since that's what CodeHelp took. This made it easier to fully explain my problem, making the answer better.'' -- NES}

\paragraph{Guardrail Appreciation}
Students mentioned how they appreciated that the answer was not explicitly given, rather they received guidance to find the answer on their own. \textit{``I appreciated how it helped me get to the answer without outright giving me the answer.'' -- NNES}

\paragraph{CodeHelp for Smaller Questions}
Students mentioned how they would typically use CodeHelp for smaller questions while reaching out to teaching assistants for conceptual questions. \textit{``If it's a relatively small problem I'll absolutely use codehelp, but for questions about the assignment or things discussed in class, I prefer to ask for help from TAs just because I feel like they are ready to answer those questions in an in depth level.'' -- NES}

In the same vein, they believed that CodeHelp worked for smaller problems due to not being able to ask questions to the system that might span across multiple files in their programming assignments. \textit{``I think that some questions I'm stuck with can't be answered because it feels like I would need to upload multiple files/long treks of code in order to fully explain my problem.'' -- NES}

\paragraph{CodeHelp Visibility and Usage Issues}
\label{subsubsubsec:results:llm-views:general:codehelp-visibility-usage-issues}
Students had mentioned that they had forgotten about CodeHelp as an available resource. \textit{``I completely forgot about this resource.'' -- NES}

Additionally, students had mentioned difficulty using the system. Possibly due to the fact that neither an in-class demo nor video of how to use CodeHelp were provided. \textit{``Perhaps a starter video on how to navigate the tool.'' -- NES}

\paragraph{NNES Difficulty Translating Terminology to Native Language}
Students mentioned interacting with the CodeHelp system in English due to learning computing in an English context. \textit{``It is doing fine with my native, but I rather do English since I learned all the course materials in English and have 0 ideas how each terms translate into my native [language].'' -- NNES}

\subsubsection{\textbf{Help Seeking Preferences}}
\label{subsubsec:results:llm-views:help-seeking}
\begin{figure*}
    \centering

    \caption{Student Preferences for Help Resources}


    \begin{subfigure}{0.75\textwidth}  
        \centering
        \caption{Native English Speakers}
        \includegraphics[width=\linewidth]{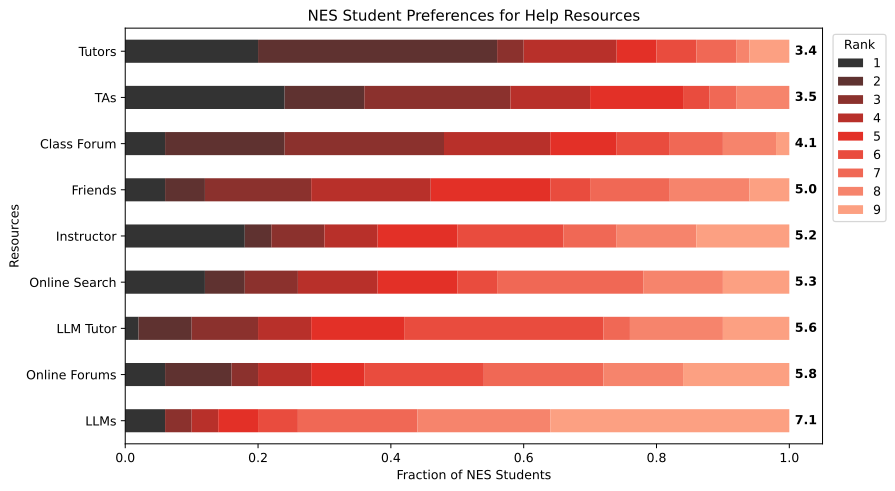}
        \label{fig:nes-preferences-fractional}
    \end{subfigure}  
    \begin{subfigure}{0.75\textwidth}  
        \centering
        \caption{Non-native English Speakers}
        \includegraphics[width=\linewidth]{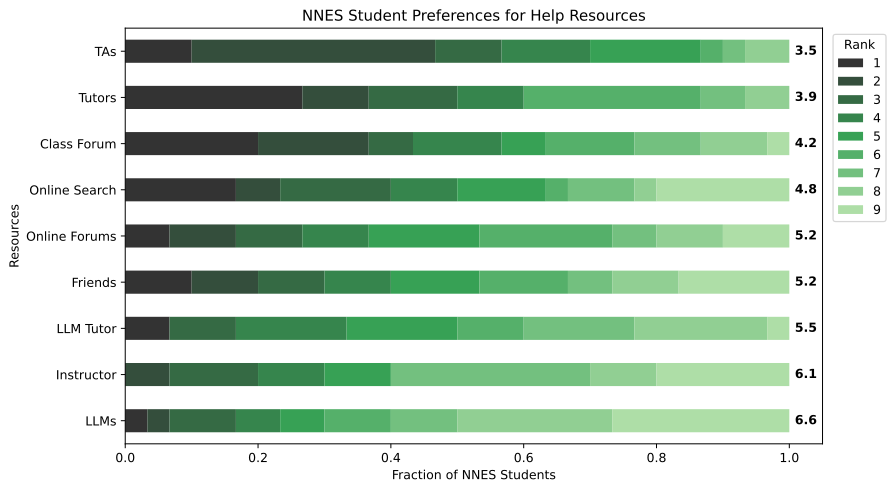}
        \label{fig:nnes-preferences-fractional}
    \end{subfigure}

    \label{fig:student-preferences-fractional}

\end{figure*}

Figure~\ref{fig:student-preferences-fractional} illustrates the results of the first survey's question exploring students' preferred approaches for help seeking.
Resources are sorted in order of the preferences of each group. The weighted averages for each resource is listed at the end of each bar where a lower average (i.e. preferred ranking) indicates a higher preference. NES preferences (Figure~\ref{fig:nes-preferences-fractional}) shows in-class and in-person resources all preferred before using online resources. NNES preferences (Figure~\ref{fig:nnes-preferences-fractional}) shows that they leverage outside resources more quickly than NES---notably leaving the instructor to the penultimate position.

\subsubsection{\textbf{LLM Tutor Evaluation}}
\label{subsubsec:results:llm-views:codehelp-evaluation}

\begin{table*}[]
\caption{Replicated Likert Questions from \citeauthor{liffiton2023codehelp}~\cite{liffiton2023codehelp}}
\label{tab:codehelp-replicated-separated-results}
\vspace{-8pt}
\begin{tabular}{L{6cm}|C{1cm}|C{1.5cm}|C{1.5cm}|C{1.5cm}|C{1.5cm}|C{1.5cm}}
\toprule
\hline
\textbf{Statement}                                                                            & \textbf{Group} & \textbf{Strongly Disagree} & \textbf{Disagree} & \textbf{Neutral} & \textbf{Agree} & \textbf{Strongly Agree} \\ \hline
\multirow{2}{6cm}{CodeHelp helped me complete my work successfully.}                                                  & NES                  & 1 (2.0\%)                  & 4 (8.0\%)         & 14 (28.0\%)      & 21 (42.0\%)    & 3 (6.0\%)                                    \\
                                                                                                                    & NNES                 & 1 (3.3\%)                  & 2 (6.7\%)         & 8 (26.7\%)       & 15 (50.0\%)    & 2 (6.7\%)                                    \\ \hline
\multirow{2}{6cm}{CodeHelp helped me learn the course material.}                                                      & NES                  & 1 (2.0\%)                  & 6 (12.0\%)        & 9 (18.0\%)       & 21 (42.0\%)    & 6 (12.0\%)                                   \\
                                                                                                                    & NNES                 & 1 (3.3\%)                  & 3 (10.0\%)        & 9 (30.0\%)       & 11 (36.7\%)    & 4 (13.3\%)                                   \\ \hline
\multirow{2}{6cm}{If I took more CS courses, I would like to be able to use CodeHelp in those classes.} & NES                  & 1 (2.0\%)                  & 1 (2.0\%)         & 4 (8.0\%)        & 16 (32.0\%)    & 21 (42.0\%)                                  \\
                                                                                                                    & NNES                 & 0 (0.0\%)                  & 2 (6.7\%)         & 4 (13.3\%)       & 8 (26.7\%)     & 14 (46.7\%)                      \\ \hline
\bottomrule
\end{tabular}
\end{table*}

We replicated the Likert questions posed to participants in the original CodeHelp study by Liffiton et al.~\cite{liffiton2023codehelp} in our first survey. Our results can be found in Table~\ref{tab:codehelp-replicated-separated-results}. We found similar results with students typically finding CodeHelp useful to complete work and learn course material. Students were largely supportive of CodeHelp being available in future CS courses. To identify significant differences, we ran an independent samples t-test across all questions comparing NNES and NES responses. There were no significant differences for any question.

\subsubsection{\textbf{LLM Tutor Views}}
\label{subsubsec:results:llm-views:survey2}

\begin{table*}
    \centering
    \caption{Likert Questions in Survey 2 Compared Across NNES and NES}
    \label{tab:survey2-questions-results}
    \vspace{-8pt}
    \begin{tabular}{L{11.25cm}|L{1cm}|L{1cm}|L{1.2cm}|L{1.5cm}}
		\toprule
		\hline
		\textbf{Question} & \textbf{NES Mean} & \textbf{NNES Mean} & \textbf{P-Value} & \textbf{Corrected P-Value} \\
		\hline
		I use CodeHelp because the professor told us we could use it in the class. & 3.80 & 3.72 & 0.6615 & 1.0000 \\ \hline
		I prefer CodeHelp to ChatGPT because it does not give me the answer directly. & 3.56 & 3.79 & 0.2056 & 0.9937 \\ \hline
		I believe that CodeHelp gives me just enough information to continue my work without providing the answer. & 3.67 & 3.77 & 0.5023 & 1.0000 \\ \hline
		I like that CodeHelp tries to guide me to an answer. & 4.06 & 3.91 & 0.3093 & 0.9997 \\ \hline
		I believe that CodeHelp assisted me in completing my work faster. & 3.49 & 3.77 & 0.0845 & 0.8567 \\ \hline
		I wish I could upload a file with my code to CodeHelp to have it help me debug all of my code for an assignment. & 3.87 & 3.87 & 0.9950 & 1.0000 \\ \hline
		I found the CodeHelp user interface easy to use. & 4.07 & 3.91 & 0.3806 & 1.0000 \\ \hline
		I find asking CodeHelp for help is easier than asking a tutor/TA. & 3.32 & 3.64 & 0.1261 & 0.9485 \\ \hline
		\textbf{I find the quality of CodeHelp responses to be better than asking a tutor.} & \textbf{2.74} & \textbf{3.13} & \textbf{0.0244} & \textbf{0.4187} \\ \hline
		I like CodeHelp for the ease of being able to ask a question. & 4.06 & 4.11 & 0.7442 & 1.0000 \\ \hline
		I like CodeHelp for the accuracy of its answers. & 3.64 & 3.74 & 0.5141 & 1.0000 \\ \hline
		I would decide when to use a digital TA (e.g., CodeHelp) or a tutor/TA based on the complexity of my question. & 3.90 & 3.75 & 0.3408 & 0.9999 \\ \hline
		I would use a digital TA (e.g., CodeHelp) more if it maintained a chat history similar to the way ChatGPT does. & 3.63 & 3.84 & 0.1795 & 0.9871 \\ \hline
		I trust the help provided by a digital TA tool (e.g., CodeHelp). & 3.57 & 3.47 & 0.5142 & 1.0000 \\ \hline
		I believe that a digital TA tool (e.g., CodeHelp) should completely help me debug my program. & 3.12 & 3.37 & 0.1761 & 0.9859 \\ \hline
		I believe that the availability of tutors/TAs/instructors is large enough to not need a digital TA tool (e.g., CodeHelp). & 2.99 & 3.26 & 0.1242 & 0.9460 \\ \hline
		I believe that if digital TA (e.g., CodeHelp) was integrated through official course platforms (e.g., Canvas, EdStem, course website, etc.) that I would use it more often. & 4.07 & 3.86 & 0.2135 & 0.9949 \\ \hline
		I wish a digital TA tool (e.g., CodeHelp) debugged my code and told me what to fix. & 3.81 & 3.67 & 0.3421 & 0.9999 \\ \hline
		I believe that I should use other sources of help rather than relying on a digital TA tool (e.g., CodeHelp). & 3.24 & 3.32 & 0.6478 & 1.0000 \\ \hline
		I feel embarrassed about bringing basic questions to a tutor/TA. & 3.33 & 3.25 & 0.6724 & 1.0000 \\ \hline
		I feel embarrassed about bringing basic questions to a digital TA tool (e.g., CodeHelp). & 2.05 & 2.19 & 0.4650 & 1.0000 \\ \hline
		I am fearful of using ChatGPT because I don't know if it's allowed to be used in this class. & 3.77 & 3.67 & 0.5700 & 1.0000 \\ \hline
		\bottomrule
    \end{tabular}
\end{table*}
Using the open-ended responses from Survey 1, we created targeted questions for Survey 2. We ran independent samples t-tests on all questions, 
using the Šídák correction to adjust the significance level to account for multiple comparisons. All questions, along with their means, p-values, and corrected p-values, are listed in Table~\ref{tab:survey2-questions-results}.
Again, no significant differences were found between NNES and NES responses for any question (post-correction).
In the initial independent samples t-tests, it was found that one question showed a significant difference between NNES and NES students at $p=0.02$ (``I find the quality of CodeHelp responses to be better than asking a tutor.''). Upon further inspection with the Šídák correction, we found that none of the comparisons were significantly different.

\subsubsection{\textbf{NNES Questions}}
\label{subsubsec:results:llm-views:nnes-questions}

\begin{table*}
	\centering
	\caption{Likert Questions for NNES Students}
	\label{tab:survey2-nnes-questions-results}
        \vspace{-8pt}
	\begin{tabular}{L{7cm}|c|c|c|c|c}
		\toprule
		\hline
		\textbf{Statement} & \textbf{Strongly Disagree} & \textbf{Disagree} & \textbf{Neutral} & \textbf{Agree} & \textbf{Strongly Agree} \\ \hline
		I find it difficult to express computing ideas in my native language. & 9 (15.8\%) & 10 (17.5\%) & 8 (14.0\%) & 14 (24.6\%) & 16 (28.1\%) \\ \hline
		I find it difficult to understand computing ideas in my native language. & 8 (14.0\%) & 11 (19.3\%) & 9 (15.8\%) & 17 (29.8\%) & 12 (21.1\%) \\ \hline
		I would be more likely to ask questions in my native language if CodeHelp's response kept the technical terminology in English (i.e., refers to an "if" statement). & 9 (15.8\%) & 9 (15.8\%) & 15 (26.3\%) & 9 (15.8\%) & 15 (26.3\%) \\ \hline
		I don't know the words for programming technical terminology in my native language (e.g., the word for "inheritance" in my language). & 6 (10.5\%) & 5 (8.8\%) & 8 (14.0\%) & 16 (28.1\%) & 22 (38.6\%) \\ \hline
		\bottomrule
	\end{tabular}
\end{table*}

Specific questions for NNES were posed in an attempt to better understand their experience in computing and their LLM tutor usage. The questions and the results can be found in Table~\ref{tab:survey2-nnes-questions-results}. We found that the NNES students in our study typically found it difficult to express and understand computing ideas in their native language, with both related questions having over 50\% agreement. Their usage of CodeHelp depending on the language of the terminology is generally mixed. Lastly, we found that NNES students in our study typically do not know the programming technical terminology in their native language.

\section{Discussion}
\label{sec:discussion}




\subsection{NNES use CodeHelp less than NES}
Our analysis revealed that NNES and NES students typically enrolled in the CodeHelp system at similar rates, which aligned with our expectations. However, contrary to our expectations, NNES students generally used CodeHelp less frequently than NES students (albeit with a small effect size). This is interesting coupled with the results found in help-seeking behaviors (Section~\ref{subsubsec:results:llm-views:help-seeking}) that showed NNES students have a similar preference for LLM tutors compared to NES students.

We observed differences based on time of day with NES using it more before the Wednesday deadline for assignments. For the NNES students, there are many possible hypotheses for this finding, such as a possible tendency to use the system more to understand the assignment throughout the week than NES students. However, we are hesitant to draw any conclusions at this stage and believe this is a potentially interesting avenue for future research.

We also found that NNES students used a larger variety of languages when interacting with CodeHelp compared to NES students. This result was expected, however it was notable to see NES students also using languages other than English. Additionally, many of the queries were multilingual using English computing keywords. This is in line with the result of NNES students not knowing computing terminology in their native language (Section~\ref{subsubsec:results:llm-views:nnes-questions}).


\subsection{NNES and NES evaluate CodeHelp similarly---but NNES rank online help higher}
We saw that NNES students tended to rank online resources, such as CodeHelp, higher than NES students. This suggests that NNES students may be more proactive in seeking assistance online---possibly due to a greater need for supplementary resources to bridge language and comprehension gaps in traditional classroom settings. This communication apprehension and linguistic anxiety from NNES with others (such as friends and the instructor) is consistent with findings from past studies~\cite{sobotka2020role, pappamihiel2001moving, agarwal2022analysis}. Interestingly, we saw that both groups rated LLM tutors above LLMs. This is possibly due to having the LLM tutor ``provided'' by the course which telegraphs an endorsement by the professor for usage in the classroom (backed by finding in Section~\ref{subsubsubsec:results:llm-views:general:instructor-approved}).

Both NNES and NES students evaluated the LLM tutor similarly, consistent with the results of the original CodeHelp study~\cite{liffiton2023codehelp}. This indicates that the LLM tutor, CodeHelp, was perceived as an effective educational tool across different studies.
Additionally, students appreciated that CodeHelp had guardrails and guided to the answer---in line with previous findings on LLM tutor scaffolding~\cite{denny2024desirable}. By offering guidance instead of direct answers, it enables students to maintain autonomy in their learning.
Lastly, we see that NNES students do not seem as comfortable with computing terminology in their native languages---in line with previous research reporting NNES student difficulties in translating this terminology~\cite{alaofi2023use}. This discomfort may be attributed to the predominance of English in computer science education and industry~\cite{hanselman2008you, chistyakov2017language, igawa2017non}, resulting in limited exposure to technical terms in other languages.

\section{Limitations}
\label{sec:limitations}
One limitation of our study could be the use of GPT-3.5-Turbo rather than GPT-4o. There is a possibility that by incorporating a state-of-the-art model the experience would have been more enjoyable for students---possibly encouraging increased use. Further work should leverage state-of-the-art models to compare results with our own.

Another limitation is in the use of LLMs with datasets that could potentially be considered too large---where some natural languages may benefit from the most advanced techniques when others may not~\cite{bender2021on}. Furthermore, LLMs have been found to exhibit stereotypes and cultural biases towards certain groups of people outside of Anglophonic settings~\cite{maina2024exploring}. Future work should actively test the LLMs used in their study on whether it displays biases for particular natural languages or cultural biases and clearly state those biases.

Additionally, we acknowledge that the introduction of CodeHelp was through a class forum post. This may contribute to the system's visibility and usage issues found in Section~\ref{subsubsubsec:results:llm-views:general:codehelp-visibility-usage-issues}. We suggest future implementations of LLM tutors provide in-class demonstrations of the system highlighting how to use the tool as well as its multilingual capabilities---possibly increasing overall usage and multilingual queries. Alongside the demonstration, an onboarding video and/or text manual for the system may serve as a valuable, asynchronous resource for students with usage issues. 



\section{Conclusion}
\label{sec:conclusion}

Computer science has historically posed challenges for NNES students due to language and terminology barriers. The rise of LLMs offers potential support for NNES students, with recent implementations of LLM-powered tutors showing promising results. In this study, we deployed an LLM tutor in an accelerated introductory computing course to evaluate its efficacy for NNES students. Our findings show that NNES students signed up for the LLM tutor at a similar rate as NES but used the system slightly less. Notably, NNES students asked significantly more questions in languages other than English, often mixing in English programming keywords.

Both NNES and NES students appreciated the LLM tutor for its accessibility, conversational style, and guidance-oriented responses. NNES students found the tool particularly approachable as it didn't require perfect English. NNES rated their help-seeking preference of online resources higher than NES students. Lastly, many NNES students were unfamiliar with computing terminology in their native languages. These results suggest that LLM tutors can be a valuable resource for NNES students in computer science education, providing tailored support that enhances their learning experience and overcomes language barriers.

\section{Acknowledgements}
This work was supported in part by the NSF Grant \#2417374, UCSD Sloan Scholars Fellowship, and Gates Millennium Scholarship. Thanks to Jackelyne García-Villegas for their feedback on this paper.

\bibliographystyle{IEEEtranSN}
\bibliography{ref.bib}


\end{document}